# RELATIONSHIP BETWEEN COLOURS OF OCHRE FROM ROUSSILLON AND CONTENT OF IRON-BEARING MINERALS

S. M. Dubiel*, J. Cieślak, J. Tarasiuk and J. Nizioł
Faculty of Physics and Computer Science, AGH University of Science and Technology, al. A. Mickiewicza 30, PL-30-059 Krakow, Poland
*dubiel@novell.ftj.agh.edu.pl

**Abstract**
Nine samples of ochre originating from the Sentier des Ocres near Roussillon, France, were studied with Mössbauer spectroscopy and Spectro-photo-colorimetry. The former yielded a quantitative phase analysis of iron containing minerals (goethite, hematite and kaolin), and the latter enabled determination of the CIE-$L^*a^*b^*$ colorimetric coefficients. Based on the results obtained it is shown that both $a^*$ and $b^*$ coordinates are responsible for the colour of the investigated ochres.

## 1. INTRODUCTION

Ochres are earthy minerals having a wide spectrum of colours that spans from pale yellow to dark red or purple. For this reason and because they are very stable, non-fading, non-bending, have a strong tinting strength and reproducible shades, they have been used as natural pigments since ancient times [1,2]. Their colour was suggested to be determined by [3,4]: (a) the presence of hematite and/or goethite as chromophore for red and yellow, respectively, (b) the exact chemical composition, and (c) the range of particle sizes. According to a recent study, the positive $a^*$ coordinate in the CIE-$L^*a^*b^*$ space is the only relevant colorimetric parameter to quantitatively characterize the colour of the ochres [5]. In this paper we give evidence that also $b^*$ coordinate is correlated with the colour, at least, in the case of the ochres from Roussillon, France, the subject of the present study.

## 2. EXPERIMENTAL

Samples were collected in field within the Sentier des Ocres place near Roussillon, France. In order to study a possible dependence of their colour on size of grains, each of the nine samples, having visually different colour, was separated into four fractions with different grain sizes viz. (1) < 20 μm, (2) 20-50 μm, (3) 50-100 μm, and (4) > 100 μm. The separation process was carried out using a wet method i.e. the investigated sample has been washed down through a nest of laboratory sieves with the finest sieve at the bottom. Respective size fractions were dried at low temperature ($60^oC$) and weighed.
Most of the investigated samples were also classified according to the Munsell colour system - the colour space that specifies colours based on three colour dimensions: hue, value (lightness), and chroma (colour purity or colourfulness) – see Table 1.

**Table** 1 A list of the investigated samples that have been classified with Munsell symbols and corresponding colours.

| Sample #no | Grain size [μm] | Munsell symbol | Colour |
|---|---|---|---|
| 1 | All fractions | 2.5YR8/4 | Pale yellow |
| 2 | All fractions | 5YR7/6 | Reddish yellow |
| 3 | 20-50, 50-100 | 10YR7/4 | Very pale brown |
| 4 | < 20 | 10YR6/6 | Brownish yellow |
| 5 | Not classified | | |
| 6 | < 20 | 7.5YR7/4 | Pink |
| 6 | Other fractions | 7.5YR6/4 | Light brown |
| 7 | < 20 | 2.5YR5/6 | Red |
| 7 | 20-50, 50-100 | 5YR5/4 | Yellowish red |
| 8 | < 20 | 5YR6/8 | Reddish yellow |
| 9 | < 20 | 2.5YR6/8 | Light red |
| 9a | 50-100 | 2.5YR5/8 | Red |
| 9b | >100 | 2.5YR5/6 | Red |

Since the quantitative phase analysis with respect to those containing iron was done in this study by means of $^{57}$Fe-site Mössbauer spectroscopy (MS), the finest fraction was chosen for the present experiment, because, as it had been revealed – see text below - it had the highest iron concentration. The exception was the sample # 9 for which three fractions, as specified in Table 1, were investigated.

Mössbauer spectra were recorded at room temperature in a transmission geometry using a standard spectrometer with a drive working in a sinusoidal mode. Gamma rays of 14.4 keV energy were supplied by a Co/Rh source of ~50 mCi activity. The investigated samples were in form of powder and had a mass of 200 mg.

Typical examples of the recorded spectra on various samples are shown in Fig. 1, while in Fig. 2 the spectra measured on different fractions of sample # 9 are displayed. It is evident from the latter; that the content of iron strongly depends on the fraction, and it significantly (as 10:5:1) decreases with the increase of the fraction’s size. The spectrum recorded on this sample for the largest fraction I. e. > 100 μm hardly contained iron, though optically the fraction looked like other fractions. In fact, as it was evidenced by the X-ray diffraction experiment, the grains of this fraction were mostly composed of quartz. On the other hand, the hyperfine field distribution curves give evidence that the relative amount of goethite and hematite hardly depends on the fraction. That, in turn, means that for the different concentration of iron in various fractions is mainly responsible quartz.

Determination of the colour coefficients $L^*a^*b^*$ was based on the measured reflectance spectrum for each sample. For that purpose the sample powder was homogenously spread and smoothed on a light absorbing substrate. The layer of the sample was enough thick to assume no influence of the substrate on the reflected light. The common end of Y-shaped waveguide that consisted of braided together one core fiber surrounded by six outer fibers was approached to the sample’s surface. To the two other ends, core and outer fibers go separately. The core fiber was attached to the light source (QTH lamp with IR part filtered off), the outer fibers collected the reflected light and led it to the spectrophotometer (Oriel MS125 coupled with Andor DV420-OE CCD camera).

The recorded light spectrum corresponds to a human vision sensitivity ranging from 380 nm to 720 nm. Taking into account a small size of the studied powder grains in comparison to the fibers diameter, it was assumed the standard cosine spatial distribution of the reflected light to hold. The reflectance was calculated as a ratio of intensities of the reflected and the incidence light beams. The incidence light intensity was determined with help of a calibrated standard reflecting surface.

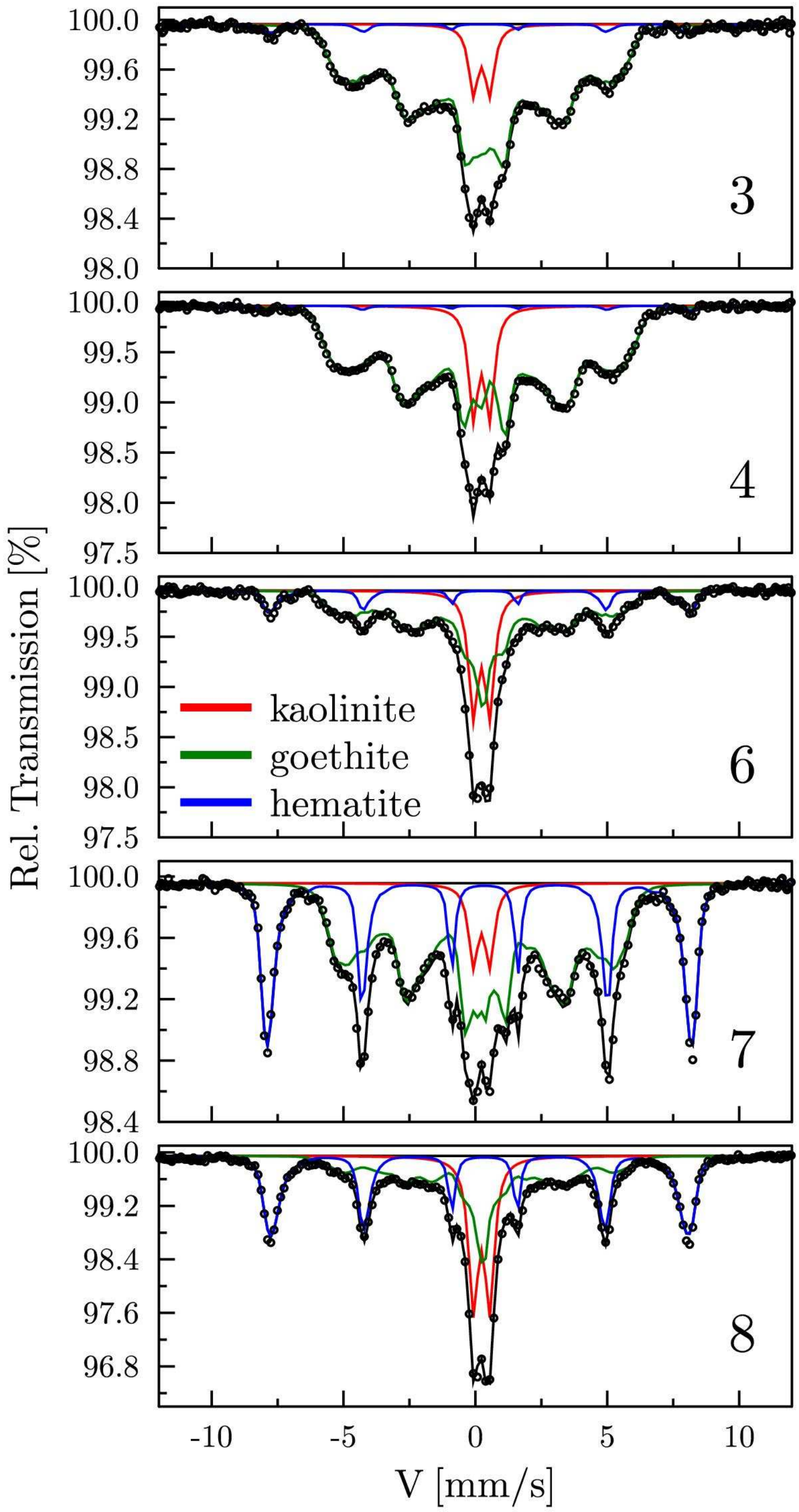

Fig. 1 $^{57}Fe$ Mössbauer spectra recorded at 295 K on the finest fraction of the ochre samples labeled with their numbers as displayed in Table 1.

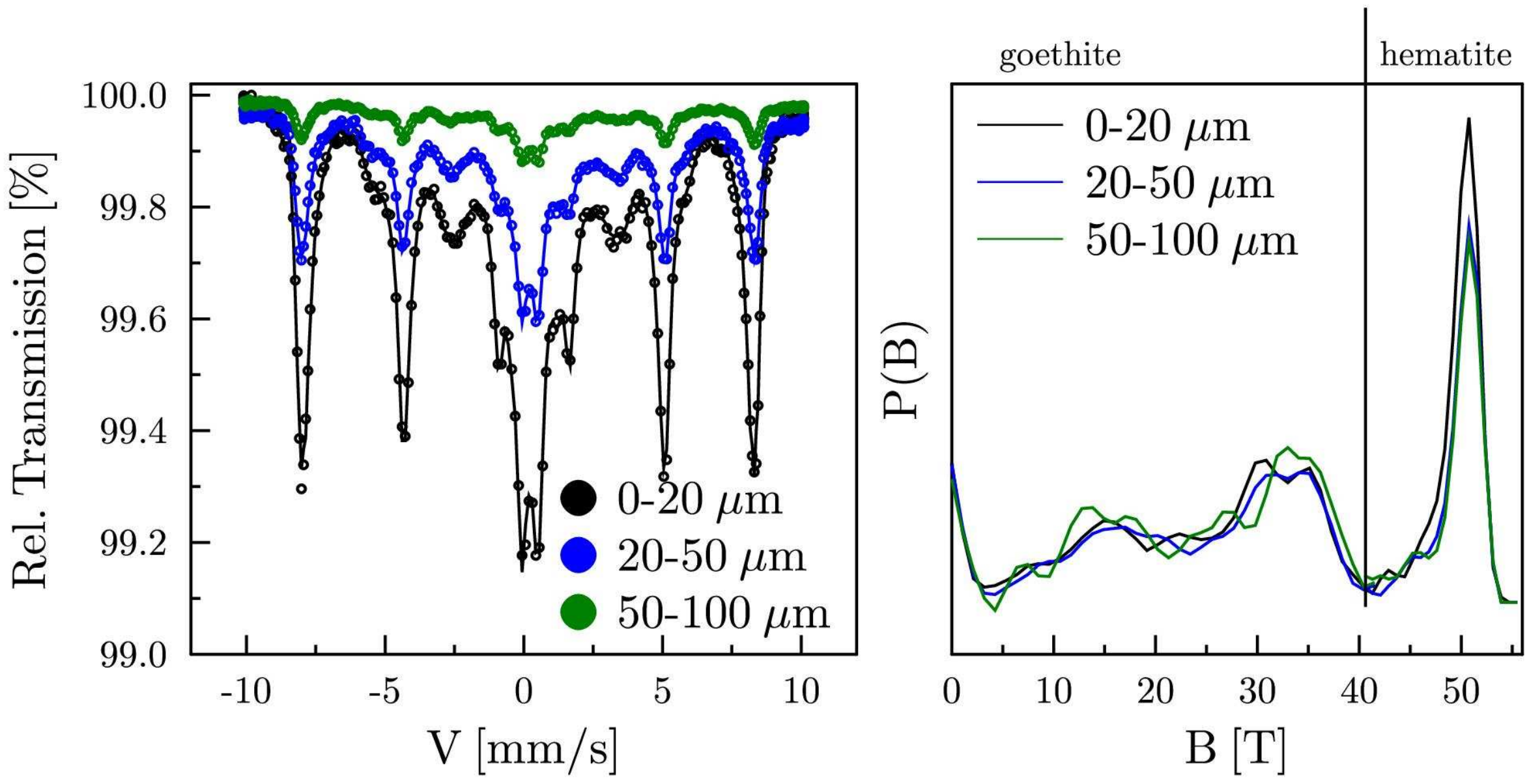

Fig. 2 $^{57}$Fe Mössbauer spectra recorded at 295 K on three different grain size fractions of the ochre's sample # 9 (left), and the corresponding distributions of the hyperfine field (right).

## 3. RESULTS AND DISCUSSION

### 3.1. Determination of iron phases from Mössbauer spectra

The phase analysis of the Mössbauer spectra was based on the assumption that the major iron-bearing minerals in the investigated samples are hematite, goethite and kaolin [3-6]. Using further the knowledge available in the literature e. g. [6,7], the hematite and goethite phases were in the fitting procedure represented by two different distributions of the hyperfine field (HFD) while the kaolin was described by a doublet. A relative amount of each phase was then determined from the relative spectral area of the corresponding spectral component. In other words, a possible effect of the Lamb-Mössbauer factor, *f*, was neglected. According to De Grave and Alboom the room temperature *f*-factor ratio between natural hematite and goethite lies in the range of 1.03 – 1.06 [8]. Since a typical error in determining spectral area from the Mössbauer spectra is of the order of 5%, then an eventual correction for the *f*-factor effect would be, in present case, meaningless.

**Table 2** Results of the phase analysis of the studied samples by means of MS in terms of kaolin, K, goethite, G, and hematite, H. Values of the best-fit spectral parameters viz. isomer shift, IS, the latter relative to the source, average hyperfine field, <B>, and quadrupole splitting, QS, are displayed, too.

| | Abundance [%] | | | IS [mm/s] | | | <B> [kOe] | | QS [mm/s] | | |
|---|---|---|---|---|---|---|---|---|---|---|---|
| # | K | G | H | K | G | H | G | H | K | G | H |
| 1 | 23.3 | 76.7 | - | 0.2345 | 0.250 | - | 216.0 | - | 0.31 | 0.12 | - |
| 2 | 15.7 | 41.1 | 43.2 | 0.2345 | 0.253 | 0.244 | 216.5 | 498.7 | 0.31 | 0.09 | 0.09 |
| 3 | 8.3 | 88.5 | 3.2 | 0.2345 | 0.250 | 0.250 | 209.0 | 497.8 | 0.31 | 0.11 | 0.10 |
| 4 | 12.9 | 85.6 | 1.5 | 0.2345 | 0.250 | 0.250 | 237.3 | 497.9 | 0.31 | 0.09 | 0.10 |
| 5 | 15.7 | 68.2 | 16.1 | 0.2345 | 0.248 | 0.246 | 217.8 | 490.6 | 0.31 | 0.11 | 0.09 |
| 6 | 23.7 | 65.4 | 10.9 | 0.2345 | 0.250 | 0.250 | 193.8 | 489.2 | 0.31 | 0.09 | 0.10 |
| 7 | 6.8 | 63.2 | 30.0 | 0.2345 | 0.251 | 0.248 | 233.2 | 494.1 | 0.31 | 0.13 | 0.10 |
| 8 | 24.1 | 38.6 | 37.3 | 0.2345 | 0.250 | 0.246 | 180.1 | 485.8 | 0.31 | 0.10 | 0.10 |
| 9 | 14.7 | 55.9 | 29.4 | 0.2345 | 0.253 | 0.246 | 216.6 | 491.4 | 0.31 | 0.09 | 0.09 |
| 9a | 17.8 | 58.3 | 23.9 | 0.2345 | 0.250 | 0.242 | 219.3 | 491.2 | 0.31 | 0.11 | 0.10 |
| 9b | 10.0 | 72.3 | 17.7 | 0.2345 | 0.250 | 0.250 | 191.2 | 489.2 | 0.31 | 0.09 | 0.10 |

### 3.2. Determination of colour coefficients

Determinations of colour coefficients have been made on the base of the measured reflectance spectrum for each sample. First *XYZ* components according to the model CIE 1964 have been calculated, using the following equations:

$$X = K\int S(\lambda)\bar{x}(\lambda)R(\lambda)d\lambda \qquad (1)$$

$$Y = K\int S(\lambda)y(\lambda)R(\lambda)d\lambda \qquad (2)$$

$$Z = K\int S(\lambda)z(\lambda)R(\lambda)d\lambda \qquad (3)$$

$$K = \frac{100}{\int S(\lambda)\bar{x}(\lambda)R(\lambda)d\lambda} \qquad (4)$$

where *x(λ), y(λ), z(λ)* are standard observer sensitivity curves (CIE 10° colour-matching functions), *S(λ)* – is intensity curve for illuminate CIE D65 (called north sky daylight) which is the standard in printing industry and is very often used in colour science analysis. Integral has been calculated in the range from 400nm to 700nm with the step of 1nm.
From *XYZ* data the *L*a*b** coefficients were determined according to following formulae:

$$L^* = 116\sqrt[3]{\frac{Y}{Y_R}} - 16 \quad \text{for} \quad \frac{Y}{Y_R} > 0.008856 \qquad (5)$$

$$L^* = 903.3\frac{Y}{Y_R} \quad \text{for} \quad \frac{Y}{Y_R} \le 0.008856 \qquad (5a)$$

$$a^* = 500\left[\sqrt[3]{\frac{X}{X_R}} - \sqrt[3]{\frac{Y}{Y_R}}\right] \qquad (6)$$

$$b^* = 200\left[\sqrt[3]{\frac{Y}{Y_R}} - \sqrt[3]{\frac{Z}{Z_R}}\right] \qquad (7)$$

where: $X$ = 94.81, $Y$ = 100.0, $Z$ = 107.3 are coordinates of reference white point for CIE D65 illuminate, $L^*$ - represents the lightness of the colour, $a^*$ - corresponds to the colour balance between red (or more accurate magenta) and green, $b^*$ - corresponds to the colour balance between yellow and green. Positive values of *a** and *b** represents warm colours (red and yellow) and negative cold colours (blue and green).

As an alternative method colour estimation based on the comparison with colour chart standard has been used. Reference QPCard 201 was applied [9]. The studied samples were photographed with reference card in the same lighting conditions. Next the software provided by the reference card producer has been used to calculate corresponding colour. The obtained colour value parameters are in a reasonable agreement with the parameters calculated on the base of the reflectance spectrum. It is important to understand that the photographic method is much less precise than the method described above and has been used only as an additional confirmation test. Consequently, for the quantitative analysis only the data obtained by the reflectance method were used. However, as can be see in Table 3, a general agreement between the results obtained with both methods was achieved. Also, as shown in Fig. 3, a fairly good correlation between colorimetric co-ordinates $a^*$ and $b^*$ obtained with the photographic method exist. The latter plot is accompanied by squares representing a comparison between the real and deduced colours for particular samples

**Table 3** Colour coefficients $a^*$, $b^*$ and $L^*$ as determined with the reflectance and photographic methods on the fraction 20-50 μm, if no specified.

| Sample No. | Reflectance method | | | Photographic method | | |
|---|---|---|---|---|---|---|
| | *L* | $a^*$ | $b^*$ | $a^*$ | $b^*$ | $L^*$ |
| 2 | 61.4 | 19.8 | 29.6 | 17 | 26 | 49 |
| 3 | 54.8 | 13.1 | 34.4 | 13 | 30 | 38 |
| 4 | 48.1 | 14.0 | 40.3 | 14 | 37 | 36 |
| 6 | 51.5 | 12.7 | 26.8 | 12 | 23 | 35 |
| 7 | 36.5 | 20.3 | 26.7 | 22 | 25 | 24 |
| 8 | 45.3 | 17.7 | 25.0 | 17 | 23 | 37 |
| $Fe_2O_3$ | 27.0 | 18.0 | 8.5 | 20 | 10 | 10 |
| $Fe_2O_3$’ all | 21.9 | 21.5 | 11.2 | 21 | 25 | 26 |
| $Fe_2O_3$, < 20 | 26.2 | 23.7 | 13.0 | No investigated | | |
| $Fe_2O_3$, 50-100 | 28.6 | 13.5 | 6.9 | | | |
| 9 | No investigated | | | 21 | 25 | 26 |
| 5 | | | | 16 | 23 | 25 |
| 1 | | | | 4 | 26 | 55 |

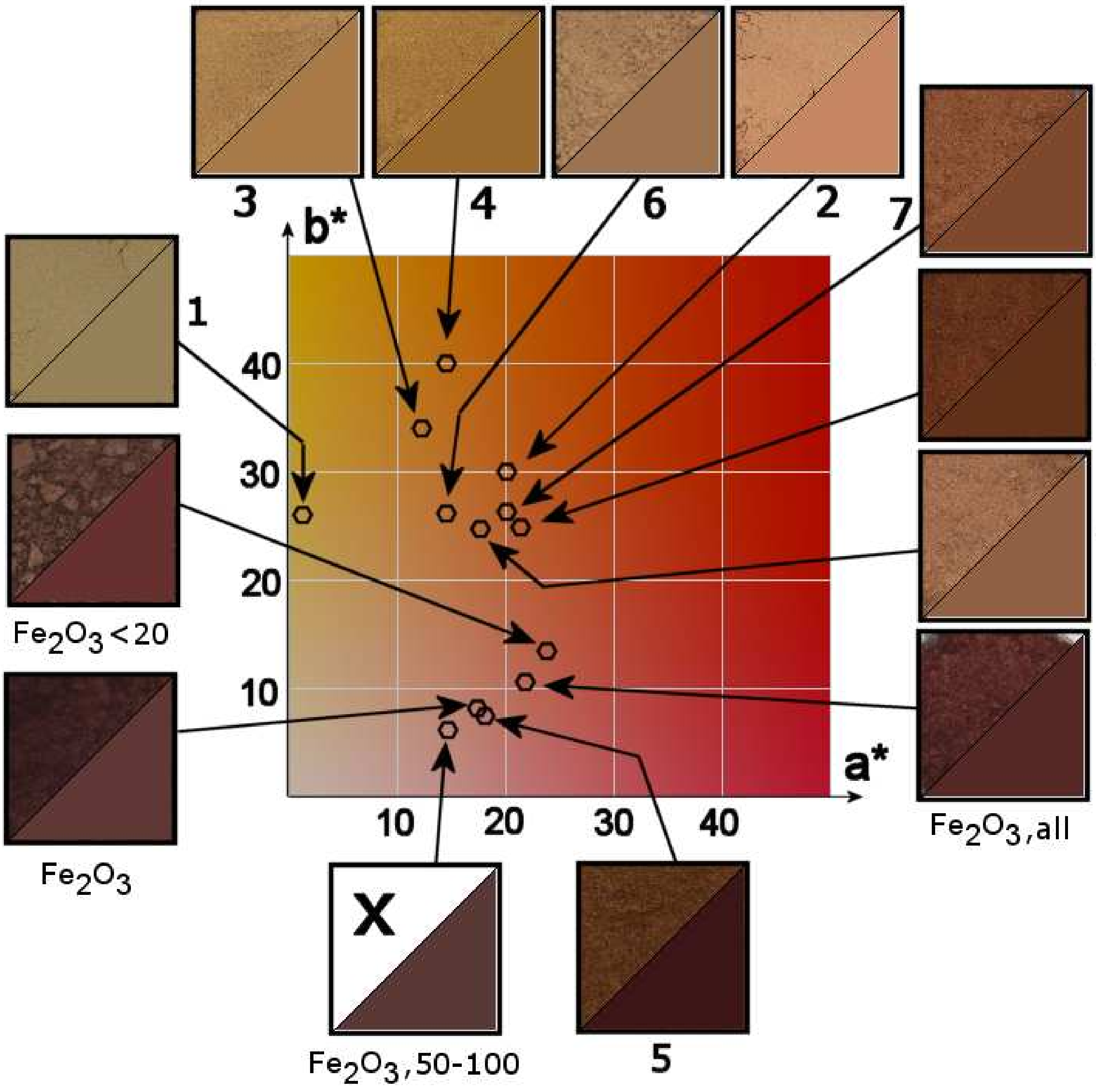

Fig. 3
Correlation between the colorimetric co-ordinates $a^*$ and $b^*$ for the indicated samples as calculated with the photographic method. The squares composed of triangles show the real (left-hand side) and calculated (right-hand side)colour of each sample (*all* is for the sample not separated into fractions, < 20 for the smallest fraction, and *x* for the non-photographed sample).

### 3.3. Discussion of the results

The colour of several samples of the ochre originating from Roussillon, France, have been quantitatively characterized in terms of the colorimetric co-ordinates in the CIE-$L^*a^*b^*$ space what has been evidenced in correlations between these coordinates as shown in Figs. 4 through 8.

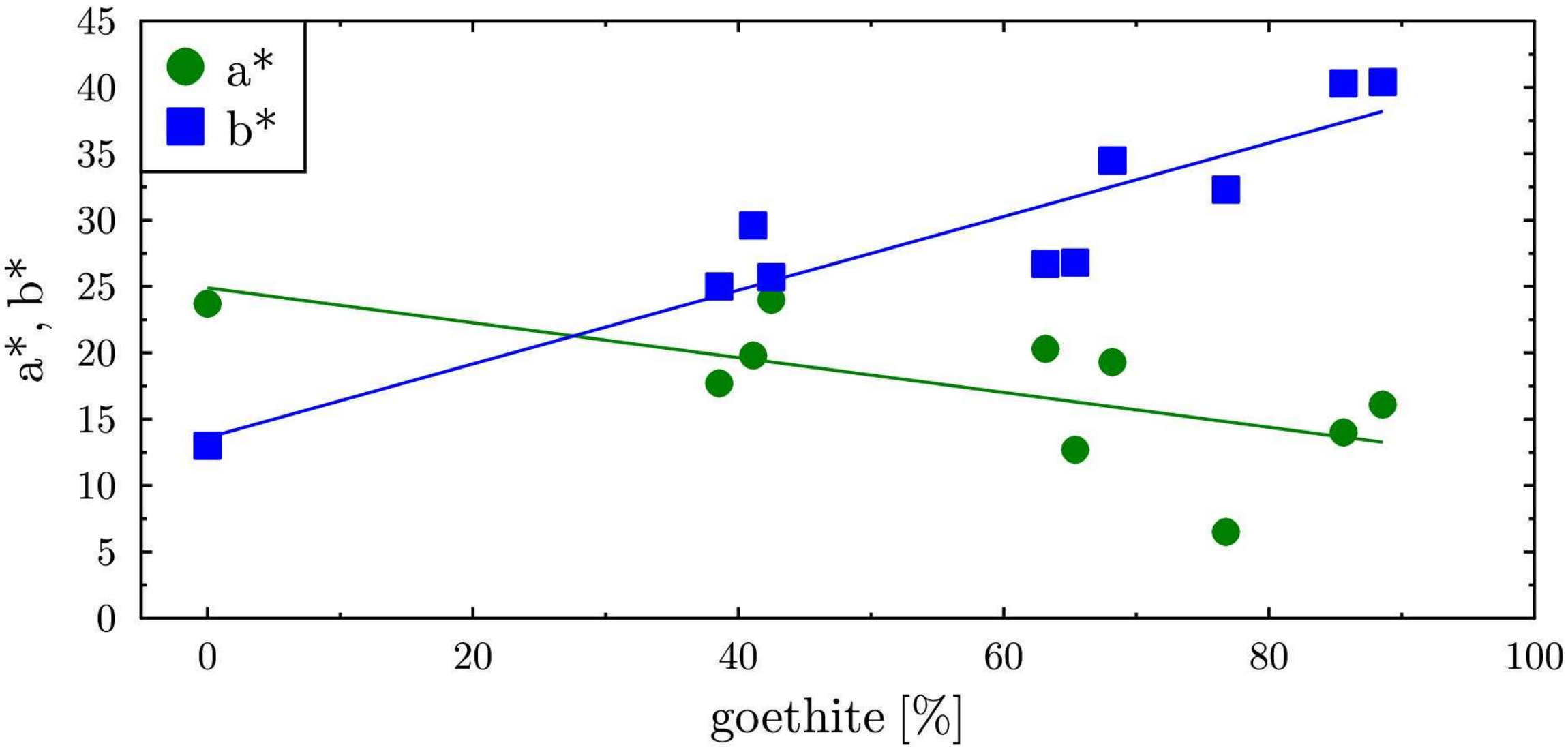

Fig. 4 Relationship between the colorimetric co-ordinates $a^*$ and $b^*$ and the content of goethite in the studied samples.

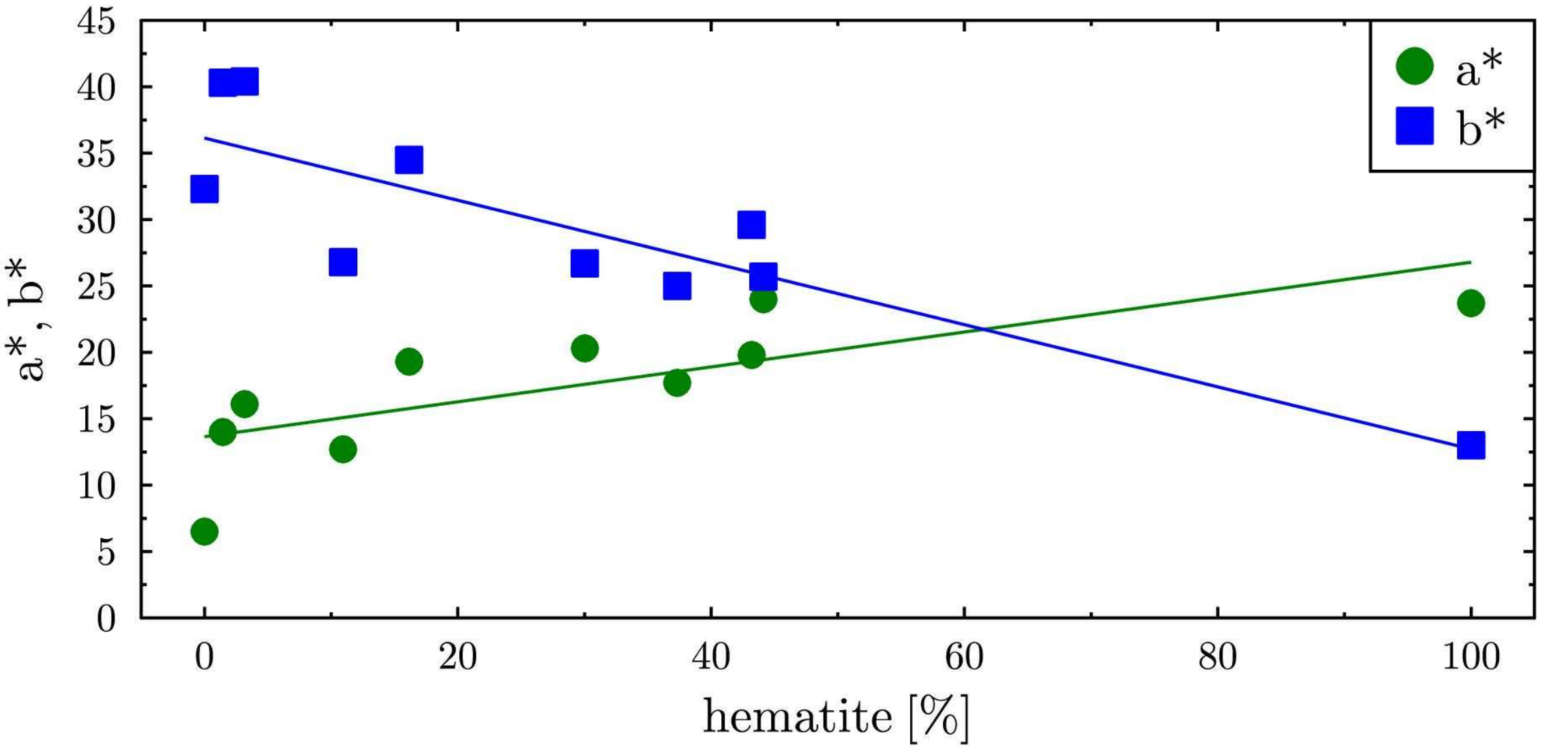

Fig. 5 Relationship between the colorimetric co-ordinates $a^*$ and $b^*$ and the content of hematite in the studied samples.

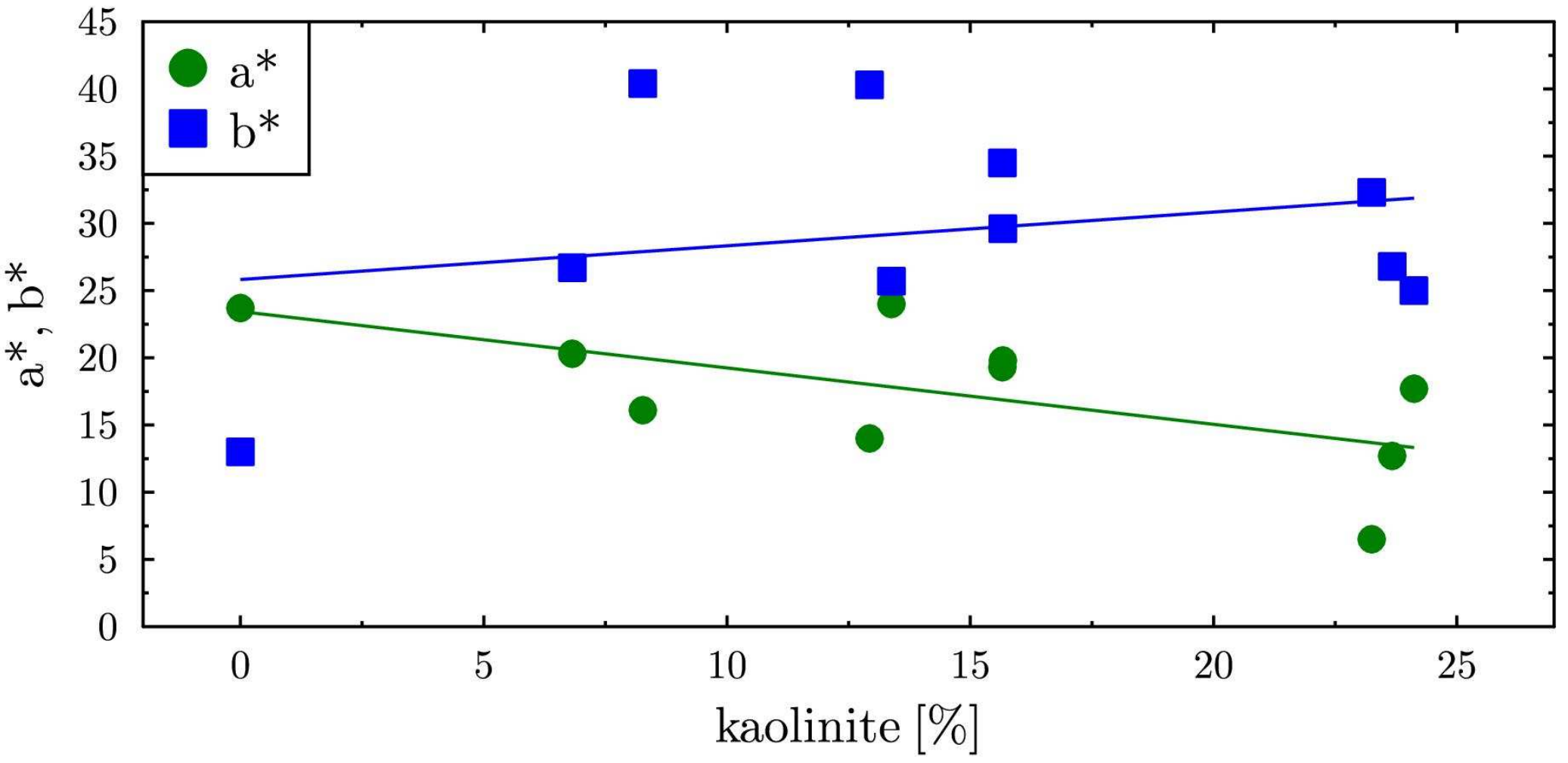

Fig. 6 Relationship between the colorimetric co-ordinates $a^*$ and $b^*$ and the content of kaolin in the studied samples.

The data shown in Figs. 4 and 5 give a clear evidence that the colorimetric co-ordinates $a^*$ and $b^*$ are linearly correlated with the content of goethite and hematite, respectively, in the studied samples. However, the correlation revealed for goethite is reversed to that found in hematite. This kind of correlations is rather expected to occur as $a^*$ corresponds to the colour balance between red (or more accurate magenta) and green, and $b^*$ is responsible for the colour balance between yellow and green. On the other hand, there is also a negative such correlation found for $a^*$ and amount of kaolin, but $b^*$ seems not to be correlated with the relative abundance of that mineral. These findings only partly agree with the previous ones [5], where only the coordinate $a^*$ was shown to be correlated with the relative amount of hematite and with that of the white pigment, hence quantitatively responsible for the colours of ochre. Here we give a clear evidence that both $a^*$ and $b^*$ play such role.

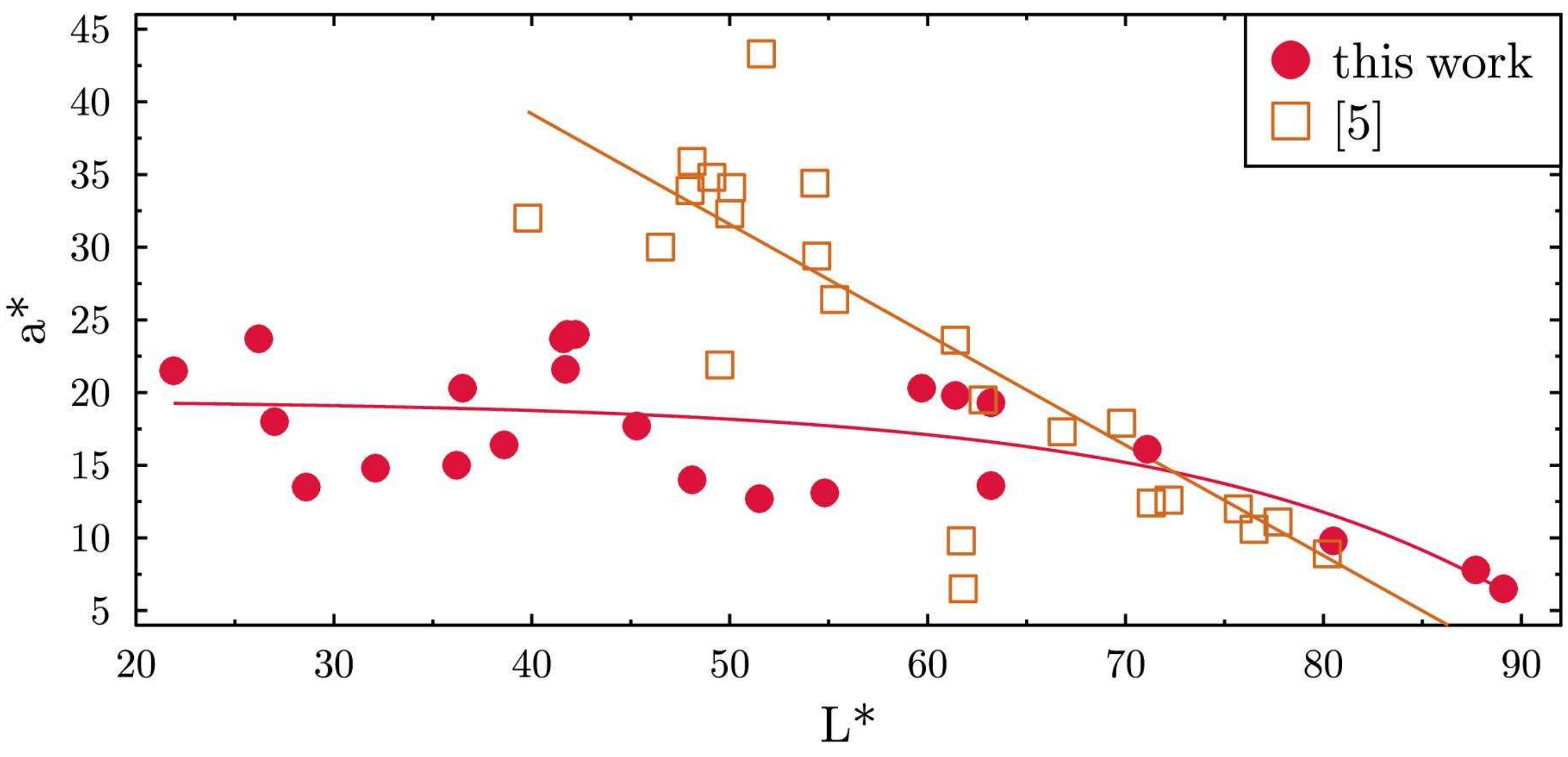

Fig. 7 Relationship between the colorimetric co-ordinates $a^*$ and $L^*$. Lines represent the best fits to the data.

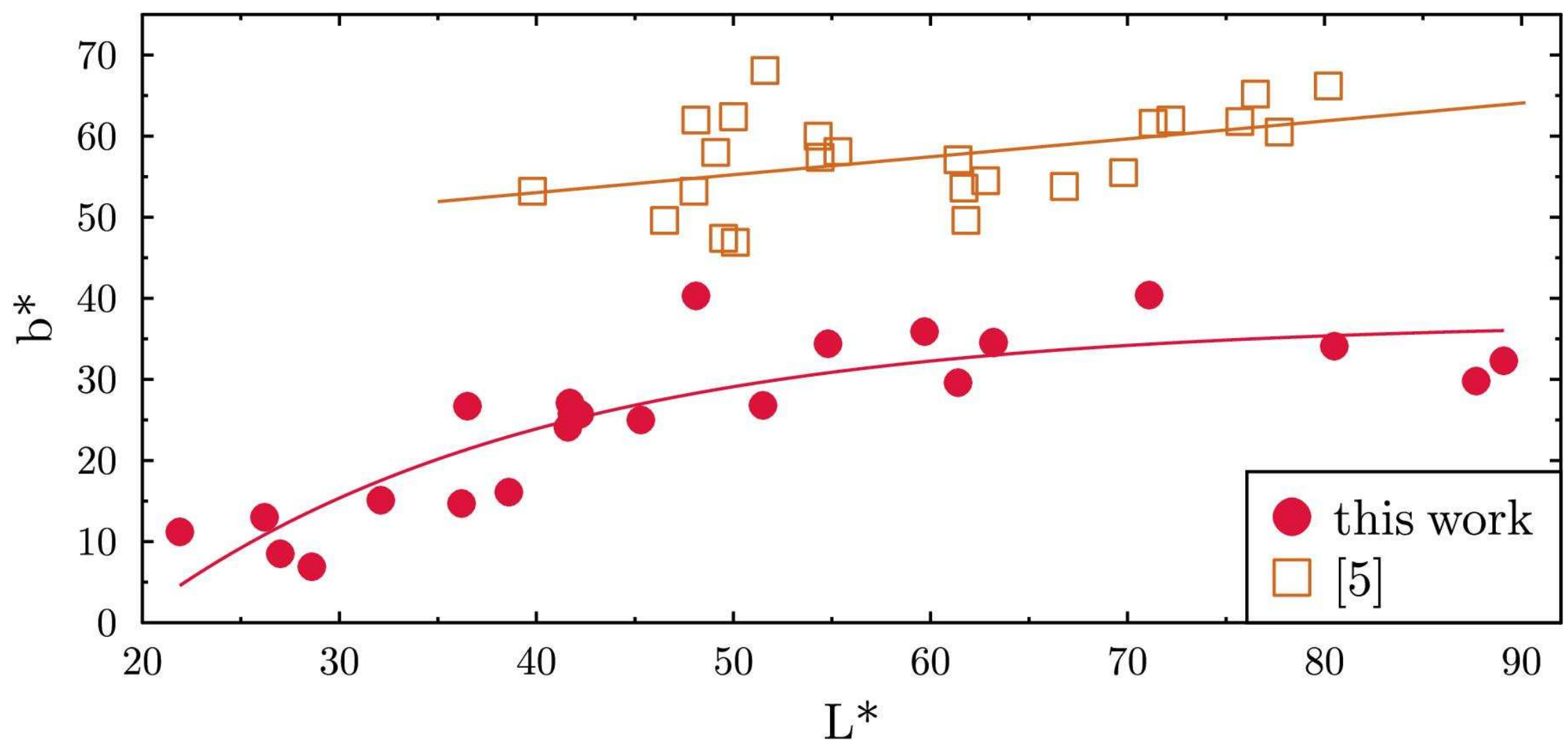

Fig. 8 Relationship between the colorimetric co-ordinates $b^*$ and $L^*$. Lines represent the best fits to the data.

As evidenced in Figs. 7 and 8, the colorimetric co-ordinates $a^*$ and $b^*$ are also correlated with $L^*$. The correlation goes in opposite directions viz. it is positive for $b^*$ and negative for $a^*$. This means that the samples of ochre with larger values of $a^*$ appear more dark, while those with larger $b^*$- values look lighter. As is evident from the two figures, the presently revealed correlations agree quantitatively with those reported elsewhere [5]. For $a^*$ and $L^*$ the presently revealed correlation is weaker than that reported in [5], and for large $L^*$ ($\geq$ ~60) it agrees quantitatively with that known from the literature [5]. For $b^*$ and $L^*$ our $b^*$ - values, that like those of $a^*$ cover much wider range of $L^*$, lie systematically below those reported elsewhere [5]. The lack of a qualitative agreement may be due both to samples of different origin as well as to various methods used by us and others to determine the phase composition of the samples. In particular, it must be here mentioned that the X-ray diffraction (XRD) method used by the authors of Ref. 5, cannot detect phases whose size of grains is lower than 1 μm that is not the case for Mössbauer spectroscopy. In other words, amount of various phases detected by using XRD may be underestimated in comparison with that found with MS.

### 3.3. Effect of size of grains

According to previous studies apparent colour of red ochre depends on size of grains [4]. Those with larger grains appear darker than those with smaller ones. To see whether or not this observation is true for our samples, and having in mind that the effect is more significant for dark red samples than for pale ones, we have carried out necessary measurements on our seemingly darkest red sample i. e. sample # 9 separated in four fractions as indicated in Fig. 9 which shows the measured diffuse reflectance spectra. Practically, no difference between them can be seen. Similar procedure was applied for a sample of the commercial hematite. The measured reflectance spectra are displayed in Fig. 10, giving evidence that in this case there is a significant difference between particular fractions. The values of $L^*$ derived from them – see Table 2 – clearly indicate that the darkness of the sample increases with the size of fraction. The discrepancy between the results obtained for the two samples may follow from the fact that our sample # 9 contains at most ~30 % of hematite i.e. it is apparently not enough dark-red to show the effect of the grain size on its colour. The data displayed in Table 2 also give a clear evidence for the sample # 9 that its actual composition is characteristic of the fraction. In particular, the relative content of hematite has the highest value for the finest fraction and it changes between 29.4 and 17.7 %, while the opposite is true for the relative concentration of goethite that has its maximum value of 72.3 % for the coarsest fraction.

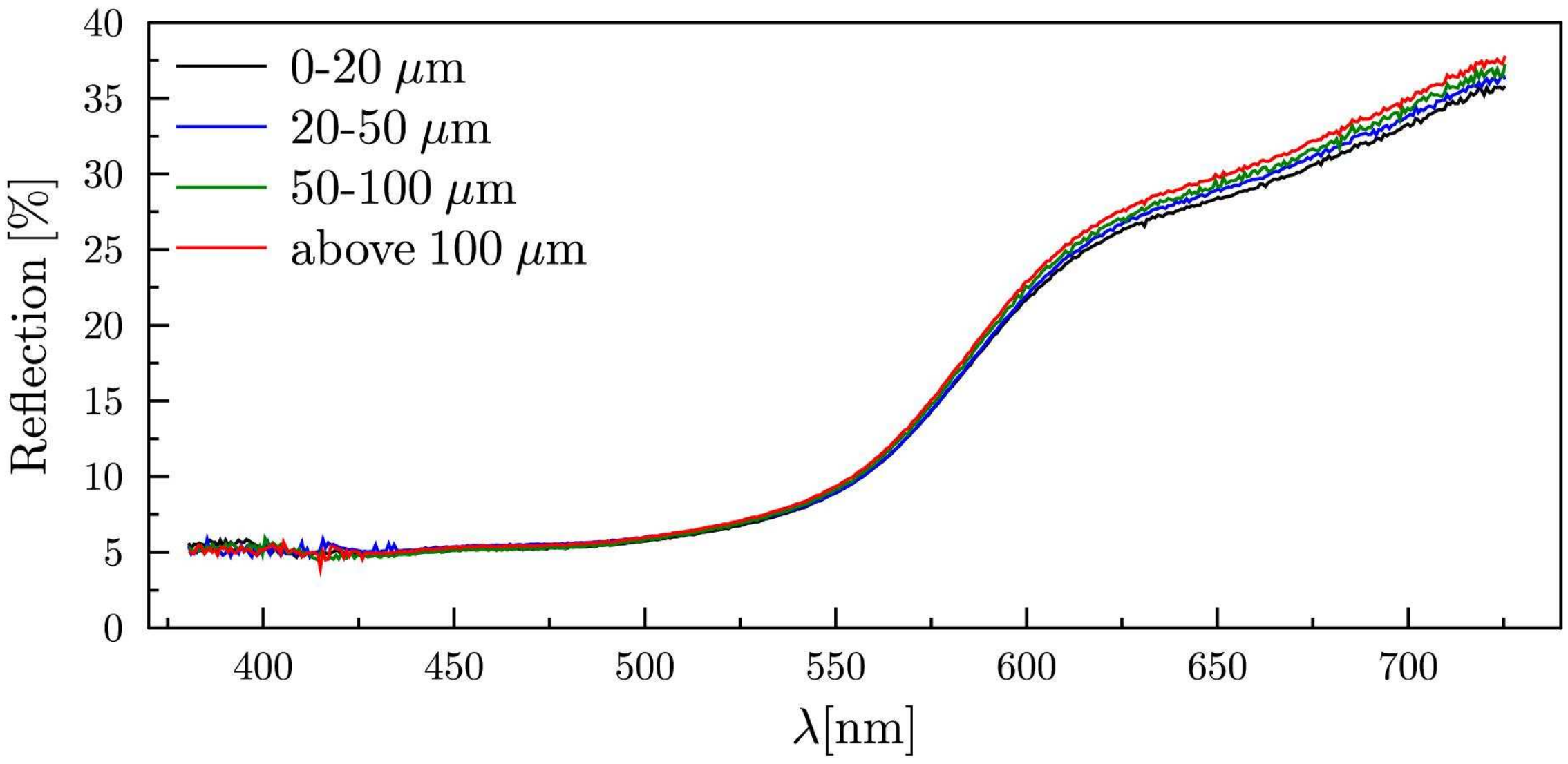

Fig. 9 Reflectance spectra measured on the sample # 9 for four different fractions.

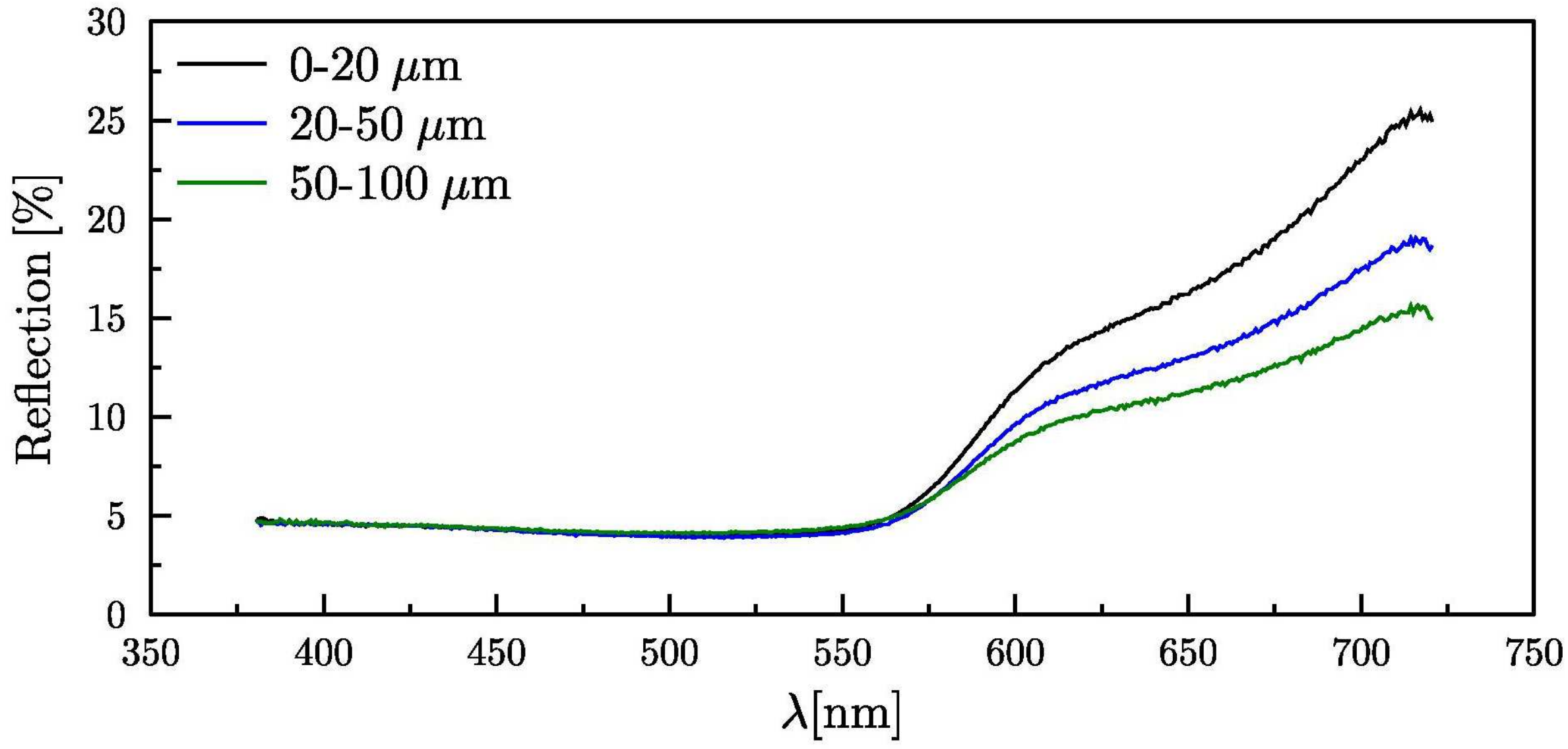

Fig. 10 Reflectance spectra measured on the sample of commercial hematite for three different fractions.

## 4. CONCLUSIONS

The colour of several samples of the ochre originating from Roussillon, France, have been quantitatively characterized in terms of the colorimetric co-ordinates in the CIE-$L^*a^*b^*$ space and compared with the amount of hematite, goethite and kaolin as found using Mössbauer spectroscopy. The results obtained can be concluded as follows:

(1) $a^*$ is positively correlated with the relative amount of hematite, negatively with that of goethite and kaolin.
(2) $b^*$ is positively correlated with the relative amount of goethite, negatively with that of hematite and hardly correlated with the amount of kaolin.
(3) $a^*$ is negatively and $b^*$ is positively correlated with $L^*$.
(4) the darkest red sample does not show any significant effect of the fraction size on its colour.

## ACKNOWLEDGEMENTS

One of us (S.M.D) wishes to thank Madame and Monsieur Deck for their support in collecting the samples. M. Brożek is acknowledged for performing separation of the samples into fractions.